\documentclass[aps,nofootinbib,showpacs,twocolumn]{revtex4-1}
\usepackage[CJKbookmarks, pdftex, bookmarksnumbered, bookmarksopen, colorlinks, citecolor=blue, linkcolor=blue]{hyperref}

\usepackage[english]{babel}
\usepackage{amstext,amsmath,amssymb,amsfonts,bbm}
\usepackage[latin1]{inputenc}
\usepackage{graphicx}
\usepackage{epstopdf}
\usepackage{amssymb}
\usepackage{bbold}
\usepackage{amsthm}
\usepackage{tocvsec2}
\usepackage{enumerate}
\usepackage{subfigure}

\usepackage{bm}
\usepackage{color}
\usepackage{stmaryrd}

\newcommand{\Ignore}[1]{}

\newcommand{\be}{\begin{equation}}
\newcommand{\ee}{\end{equation}}
\newcommand{\bes}{\begin{eqnarray}}
\newcommand{\ees}{\end{eqnarray}}
\newcommand{\Ket}[1]{\left\vert #1\right\rangle}
\newcommand{\Bra}[1]{\left\langle #1\right\vert}

\newcommand{\MV}[1]{\left\langle #1 \right\rangle}
\renewcommand{\eqref}[1]{Eq.~(\ref{#1})}

\begin{document}

\title{Typicality in spin network states of quantum geometry}

\author{Fabio Anza$^{1}$, Goffredo Chirco$^{2}$}

\affiliation{$^1$Clarendon Laboratory, University of Oxford, Parks Road, Oxford OX1 3PU, United Kingdom.\\
$^2$Max Planck Institute for Gravitational Physics, Albert Einstein Institute, Am M\"{u}hlenberg 1, 14476, Potsdam, Germany.}

\date{\today}

\begin{abstract}
In this work we extend the so-called typicality approach, originally formulated in statistical mechanics contexts, to $SU(2)$ invariant spin network states. Our results do not depend on the physical interpretation of the spin-network, however they are mainly motivated by the fact that spin-network states can describe states of quantum geometry, providing a gauge-invariant basis for the kinematical Hilbert space of several background independent approaches to quantum gravity. The first result is, by itself, the existence of a regime in which we show the emergence of a typical state. We interpret this as the prove that, in that regime there are certain (local) properties of quantum geometry which are ``universal''. Such set of properties is heralded by the typical state, of which we give the explicit form. This is our second result. In the end, we study some interesting properties of the typical state, proving that the area-law for the entropy of a surface must be satisfied at the local level, up to logarithmic corrections which we are able to bound.
\end{abstract}

\maketitle

\section{Introduction}

In recent years, quantum statistical mechanics and quantum information theory have played an increasingly central role in quantum gravity. Such interplay has proved particularly insightful both in the context of the holographic duality in AdS/CFT \cite{mal, rt, vr, sw, cz}, as well as for the current background independent approaches to quantum gravity, including loop quantum gravity (LQG)  \cite{ thi, 5, CarFra, bia}, the related   spin-foam formulation \cite{6}, and group field theory \cite{ori}. 

Interestingly, the different background independent approaches today share a microscopic description of  space-time geometry given in terms of discrete, pre-geometric degrees of freedom of combinatorial and algebraic nature, based on spin-network Hilbert spaces \cite{penrose, ro, baez1,baez2, seth}. In this context, entanglement has provided a new tool to characterise the quantum texture of space-time in terms of the structure of correlations of the spin networks states.

Along this line, several recent works have considered the possibility to use specific features of the short-range entanglement in quantum spin networks (area law, thermal behaviour) to select quantum geometry states which may eventually lead to smooth spacetime geometry classically \cite{bm, chi, chi1,ha, pra}. 
This analysis usually focuses on states with few degrees of freedom, leaving open the question of whether a \emph{statistical} characterisation may reveal new structural properties, independent from the interpretation of the spin network states.

In this work we proposes the use of the information theoretical notion of quantum \emph{canonical typicality}, as a tool to investigate and characterise universal local features of quantum geometry, going beyond the physics of states with few degrees of freedom.

In quantum statistical mechanics, \emph{canonical typicality} states that almost every pure state of a large quantum mechanical system, subject to the fixed energy constraint, is such that its reduced density matrix over a sufficiently small subsystem is approximately in the {\it canonical} state described by a thermal distribution \`{a} la Gibbs \cite{1,2a,2b,2c,3}. 

Such a statement goes beyond the thermal behaviour. 
For a generic closed system in a quantum pure state, subject to some \emph{global constraint}, the resulting canonical description will not be thermal, but generally defined in relation to the constraint considered \cite{popescu,popescu2}. Again, in this case, some specific properties of the system emerge at the local level, regardless of the nature of the global state. These properties depend on the physics encoded in the choice of the global constraints.

Within this generalised framework, we exploit the notion of \emph{typicality} to study whether and how  ``universal''  statistical features of the local correlation structure of a spin-network state emerge in connection with the  choice of the global constraint.

We focus our analysis on the space of the $N$-valent $SU(2)$ invariant intertwiners, which are the building blocks of the spin network states. In LQG, such intertwiners can be thought of dually as a region of $3d$ space with an $S^2$  boundary \cite{ro}. 

We reproduce the typicality statement in the full space of $N$-valent intertwiners with fixed total area and we investigate the statistical behaviour of the canonical reduced state, dual to a small patch of the $S^2$  boundary, in the large $N$ limit.
Eventually, we study the entropy of such a reduced state and its area scaling behaviour in different thermodynamic regimes.

The content of the manuscript is organised as follows. Section \ref{tysn} introduces the statement of canonical typicality in a formulation particularly suitable for the spin network Hilbert space description. Section \ref{snst} shortly reviews the notion of state of quantum geometry in terms of spin network basis. In Section \ref{cano} we reformulate the statement of quantum typicality in this context. We derive the notion of canonical reduced state of the $N$-valent intertwiner spin network system in Section \ref{redu} and we prove the existence of a regime of typicality for such system in Section \ref{typ}. The entropy of the typical state and its thermodynamical interpretation are investigated in Section \ref{thermo}. We conclude in Section \ref{fine} with a short discussion of our results. Technical details of all the computations are given in the Supplementary Material.

\section{Typicality}  \label{tysn}

We start with a brief summary of the result achieved in \cite{popescu}. Suppose we have a generic \emph{closed} system, which we call ``universe'', and a bipartition into ``small system'' $S$ and ``large environment'' $E$. The universe is assumed to be in a pure state. We also assume that it is subject to a completely arbitrary \emph{global constraint} $\mathcal{R}$. For example, in the standard context of statistical mechanics it can be the fixed energy constraint. Such constraint is concretely imposed   by restricting the allowed states to the subspace $\mathcal{H}_{\mathcal{R}}$ of the states of the total Hilbert space $\mathcal{H}_U$ which satisfy the constraint $\mathcal{R}$:
\begin{align}
\mathcal{H}_{\mathcal{R}} \subseteq \mathcal{H}_{U} = \mathcal{H}_E \otimes \mathcal{H}_S \,\,\, .
\end{align}
$\mathcal{H}_S$ and $\mathcal{H}_E$ are the Hilbert spaces of the system and environment, with dimensions $d_S$ and $d_E$, respectively. We also need the definition of the canonical state of the system $\Omega_S$, obtained by tracing out the environment from the microcanonical (maximally mixed) state $\mathcal{I}_{\mathcal{R}}$

\begin{align}
&\Omega_S  \equiv \mathrm{Tr}_E [\mathcal{I}_{\mathcal{R}}] &&\mathcal{I}_{\mathcal{R}} \equiv \frac{\mathbb{1}_{\mathcal{R}} }{d_{\mathcal{R}}}
\end{align}

where $\mathbb{1}_{\mathcal{R}} $ is the projector on $\mathcal{H}_{\mathcal{R}}$, and $d_{\mathcal{R}} = \mathrm{dim} \,\mathcal{H}_{\mathcal{R}}$. This corresponds to assigning {\it a priori} equal probabilities to all states of the universe consistent with the constraints ${\mathcal{R}}$.

In this setting, given an arbitrary pure state of the universe satisfying the constraint $\mathcal{R}$, i.e. $\Ket{\phi} \in \mathcal{H}_{\mathcal{R}}$, the reduced state $\rho_S(\phi) \equiv \mathrm{Tr}_E[\Ket{\phi}\Bra{\phi}]$ will {almost} always be very close to the canonical state $\Omega_S$.

Concretely, such a behaviour can be stated as a theorem \cite{popescu}, showing that for an arbitrary $\varepsilon > 0$, the distance between the reduced density matrix of the system $\rho_S(\phi)$ and the canonical state $\Omega_S$ is given probabilistically by

\begin{align}
\frac{\mathrm{Vol} \left[  \phi \in \mathcal{H}_{\mathcal{R}} \, \vert \, D(\rho_S(\phi),\Omega_S) \geq \eta \right] }{\mathrm{Vol} \left[ \phi \in \mathcal{H}_{\mathcal{R}} \right] } \leq \eta' \label{result}
\end{align}
where the {\it trace-distance} $D$ is a metric \footnote{We use the definition $D(\rho_1,\rho_2) = \frac{1}{2} \sqrt{(\rho_1 - \rho_2)^{\dagger}(\rho_1 - \rho_2)}$.} on the space of the density matrices \cite{Geos, Nielsen}, while
\begin{align}
& \eta' = 4\,  \mathrm{Exp} \left[ - \frac{2}{9\pi^3} d_{\mathcal{R}} \varepsilon^2 \right] && \eta = \varepsilon + \frac{1}{2} \sqrt{\frac{d_S}{d_E^{\mathrm{eff}}}}, 
\end{align}
with the effective dimension of the environment defined as
\begin{align}
 d_E^{\mathrm{eff}} \equiv \frac{1}{\mathrm{Tr}_E \left[ \left( \mathrm{Tr}_S \mathcal{I}_{\mathcal{R}} \right)^2\right] } \geq \frac{d_{\mathcal{R}}}{d_S} \label{deff}.
\end{align}
The bound in \eqref{result} states that the fraction of the volume of the states which are far away from the canonical state $\Omega_S$ more than $\eta$ decreases exponentially with the dimension of the ``allowed Hilbert space'' $d_{\mathcal{R}} = \mathrm{dim} \mathcal{H}_{\mathcal{R}}$ and with $\varepsilon^2 = \left(\eta - \frac{1}{2} \sqrt{\frac{d_S}{d_E^{\mathrm{eff}}}}\right)^2$. This means that, as the dimension of the Hilbert space $d_{\mathcal{R}}$ grows, a huge fraction of states gets concentrated around the canonical state. 
%

The proof of the result relies on the \emph{concentration of measure phenomenon}. The key tool to prove this result is the \emph{Levy} lemma, which we briefly report for completeness in Appendix \ref{levy}.\\

In the following we will reconsider this argument for a peculiar class of high dimensional quantum systems, the quantum spin network states. The interest in testing the notion of typicality in quantum gravity resides in the kinematic nature of the statement, a fundamental feature to study the possibility of a thermal characterisation of reduced states of quantum geometry, regardless of any hamiltonian evolution in time.

\section{states of quantum geometry} \label{snst}

In several background-independent approaches to quantum gravity, the spin network states provide a \emph{kinematical} description of quantum geometry, in terms of superpositions of graphs $\Gamma$ labelled by group or Lie algebra elements representing holonomies of the gravitational connection and their conjugate triad \cite{ro, thi,CarFra}.
 
These states are constructed as follows (for a thorough introduction to spin-networks we refer to \cite{penrose,ro,baez1,baez2,seth}). To each edge $e \in \Gamma$ one associates an $SU(2)$ irreducible representation (irrep) labelled by a half-integer $j_e \in \mathbb{N}/2$ called spin. The representation (Hilbert) space is denoted $V^{j_e}$ and has dimension $d_{j_e} = 2 j_e +1$. To each vertex $v$ of the graph one attaches an intertwiner $\mathcal{I}_v$, which is $SU(2)$-invariant map between the representation spaces $V^{j_e}$ associated to all the edges $e$ meeting at the vertex $v$,
\begin{align}
\mathcal{I}_v: \bigotimes_{e \,\,\text{ingoing}}  V^{j_e}\to \bigotimes_{e\,\, \text{outgoing}} V^{j_e}
\end{align}
One can alternatively consider $\mathcal{I}_v$ as a map from $\otimes_{e\in v} V^{j_e}\to \mathbb{C}\simeq V^0$ and call the intertwiner an invariant tensor or a singlet state between the representations attached to all the edges linked to the considered vertex. Once the $j_e$'s  are fixed, the intertwiners at the vertex $v$ actually form a Hilbert space, which we will call 
\begin{align}
\mathcal{H}_v \equiv  \text{Int}_v [ \bigotimes_{e}V^{j_e}]
\end{align}

A spin network state $|\Gamma, \{j_e\},\{\mathcal{I}_v\}\rangle$ is defined as the assignment of representation labels $j_e$ to each edge and the choice of a vector $|\{\mathcal{I}_v\}\rangle \in \otimes_v\mathcal{H}_v$ for the vertices. The spin network state defines a wave function on the space of discrete connections $SU(2)^E/SU(2)^V$ ,
\begin{align}
\phi_{\{j_e\},\{\mathcal{I}_v\}} [g_e ] = \langle h_e |  \mathcal{I}_v \rangle \equiv  \text{tr} \bigotimes_e D^{j_e}(h_e) \otimes \bigotimes_v \mathcal{I}_v
\end{align}
where we contract the intertwiners $\mathcal{I}_v$ with the (Wigner) representation matrices of the group elements $g_e$ in the chosen representations $j_e$. 

Therefore, upon choosing a basis of intertwiners for every assignment of representations ${j_e}$, the spin networks provide a basis of the space of wave functions associated to the graph $\Gamma$, 
\begin{align}
\mathcal{H}_{\Gamma} = L_2[SU(2)^E/SU(2)^V]= \bigoplus_{\{j_e\}} \bigotimes_v \mathcal{H}_v .
\end{align} 

\begin{figure}[t]
\includegraphics[width=2.5 in]{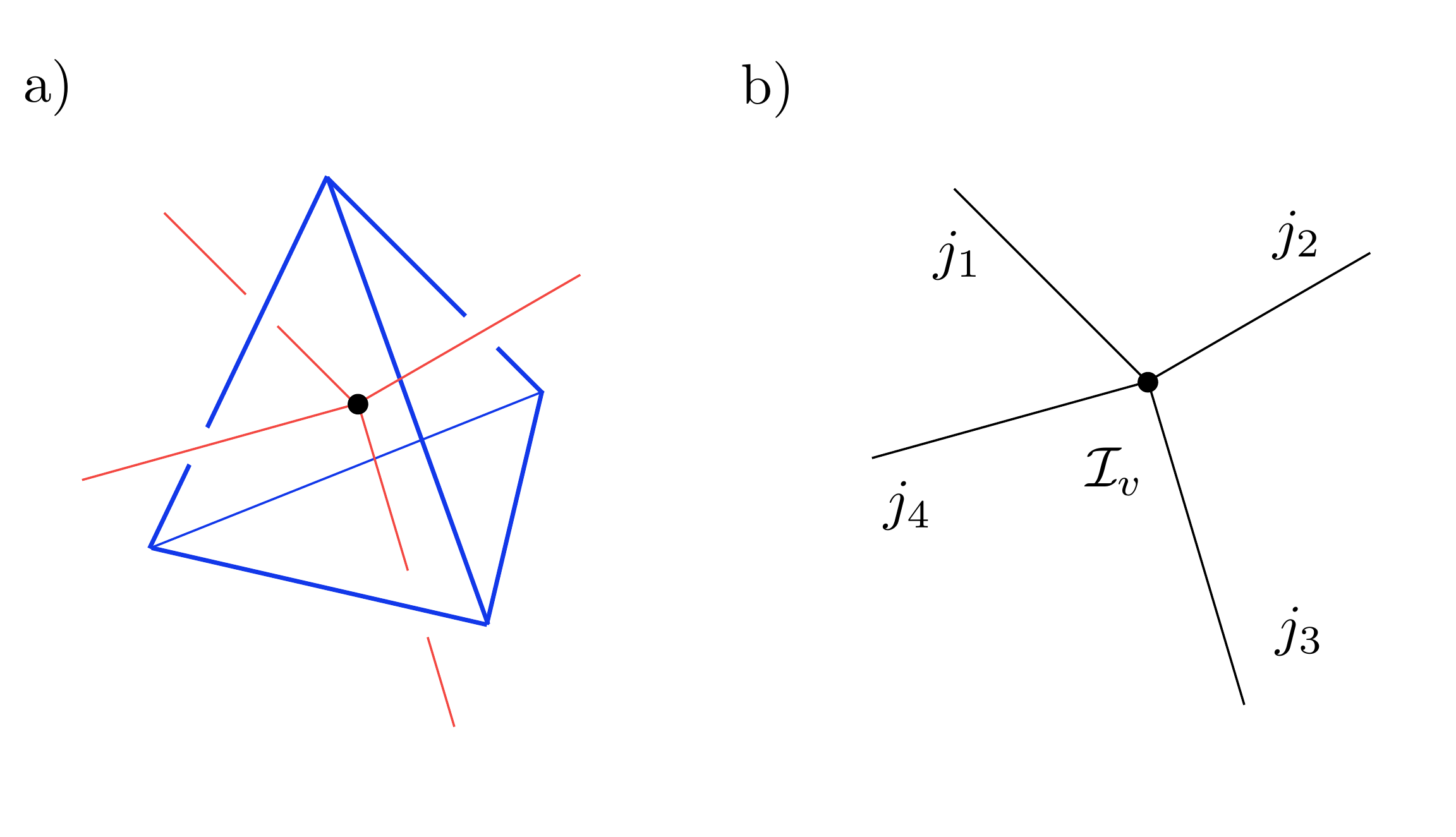}
\caption{a) Example of a four-valent node dual to a tetrahedron, describing the fundamental cell of a triangulated 3d space. b) The edges of the dual graph are labelled with spins $\{j_i\}$, while $\mathcal{I}_v$ labels the intertwiner tensor at the node.}\label{duality}
\end{figure}

Such discrete and algebraic objects provide a description of the fundamental excitations of quantum spacetime. From a geometrical point of view, classically, given a cellular decomposition of a three-dimensional manifold, a spin-network graph with a node in each cell and edges connecting nodes in neighbouring cells is said to be dual to this cellular decomposition. Therefore, each edge of the graph is dual to a surface patch intersecting the edge and the area of such patch is proportional to the representation $j_e$. Analogously, vertices of a spin network can be dually thought of as chunks of volume (see Fig.\ref{duality} for an example). See \cite{don, Poly1, Poly2, Poly3, Poly4} for the geometric interpretation of spin-networks states as collection of polyhedra.

In the following, we will focus on a fundamental building block of a spin network graph, the Hilbert space of a single intertwiner with $N$ legs.

\section{Intertwiner typicality}\label{cano}

Now we consider a large quantum system given by a collection of $N$ edges, represented by $N$ independent edges states. The Hilbert space of the system is the direct sum over $\{	j_i\}$'s of the  tensor product of $N$ irreducible representations $V^{j_i}$, 
\begin{align}\label{space}
\mathcal{H}= \bigoplus_{\{j_i\}} \bigotimes_{i=1}^N V^{j_i}.
\end{align} 
This set of independent edges plays the role of the ``universe''. 
Notice that, despite its extreme simplicity, this system has a huge Hilbert space. The single representation space $V^j$ has finite dimension $d_{j_i} = 2 j_i +1$. However, $d_{j_i}$ is summed over all $j_i \in \frac{\mathbb{N}}{2}$.  Therefore each Wilson line state (edge) lives in an infinite dimensional Hilbert space\footnote{An important detail is how we deal with spin-0 representations. In LQG these are avoided introducing  cylindrical consistency which requires that such links are equivalent to non-existent links. We do not require cylindrical consistency, hence spin-0 representations are allowed.}. In the following, we will always consider a cut-off in the value of the $SU(2)$ representation labelling the edge\footnote{Another way to introduce a cut-off in the representations, which has already been explored in literature \cite{QGroups, QuGrLQG,QuGrLQG1,QuGrLQG2,QuGrLQG3,QuGrLQG4}, is to consider the so-called q-deformation of $SU(2)$. This is usually done in LQG to include a cosmological constant.},

\begin{align}
\bigoplus_{\{j_i\}}\to \bigoplus_{\{j_i\}}^{j_i \le {J}_{max}}, \quad \text{with} \quad {J}_{max}\gg1.
\end{align}
This will allow us to deal with a very large but finite dimensional space.

Now, we want to split the universe into system and environment. We do so simply by defining two subsets of edges $E$ and $S$, with $\{1,\cdots , k \} \in {S}$  and $\{ k+1,\cdots , N \} \in{E}$, such that $k \ll N$. Consequently, we can write the Hilbert space of the universe as the tensor product $\mathcal{H}=\mathcal{H}_E \otimes \mathcal{H}_S$. We would like to stress here that, for our result to hold the $k$ links of the system do not need to be adjacent. Despite that, we are interested in the local properties of an intertwiner therefore we will always think about these links as adjacent and forming a simply connected 2D patch.\\

The next step toward typicality consists in defining the constraint which restricts the allowed states of the system and environment to a subspace of the total Hilbert space.

\subsection{Definition of the constraint}

Two main ingredients are necessary to the definition of the constraint. The first one is the $SU(2)$ gauge invariance. This choice reduces the universe Hilbert space to the collection of the $SU(2)$-{invariant} linear spaces
\begin{align}\nonumber
\mathcal{H}_{N}=\bigoplus_{\{j_i\}}\text{Inv}_{SU(2)} [\bigotimes_{i=1}^N V^{j_i}],
\end{align} 
spanned by $N$-valent intertwiner states. 
Invariance under $SU(2)$ is the the first ingredient defining our subsystem constrained space.  
\begin{figure}[t]
\includegraphics[width=2.5 in]{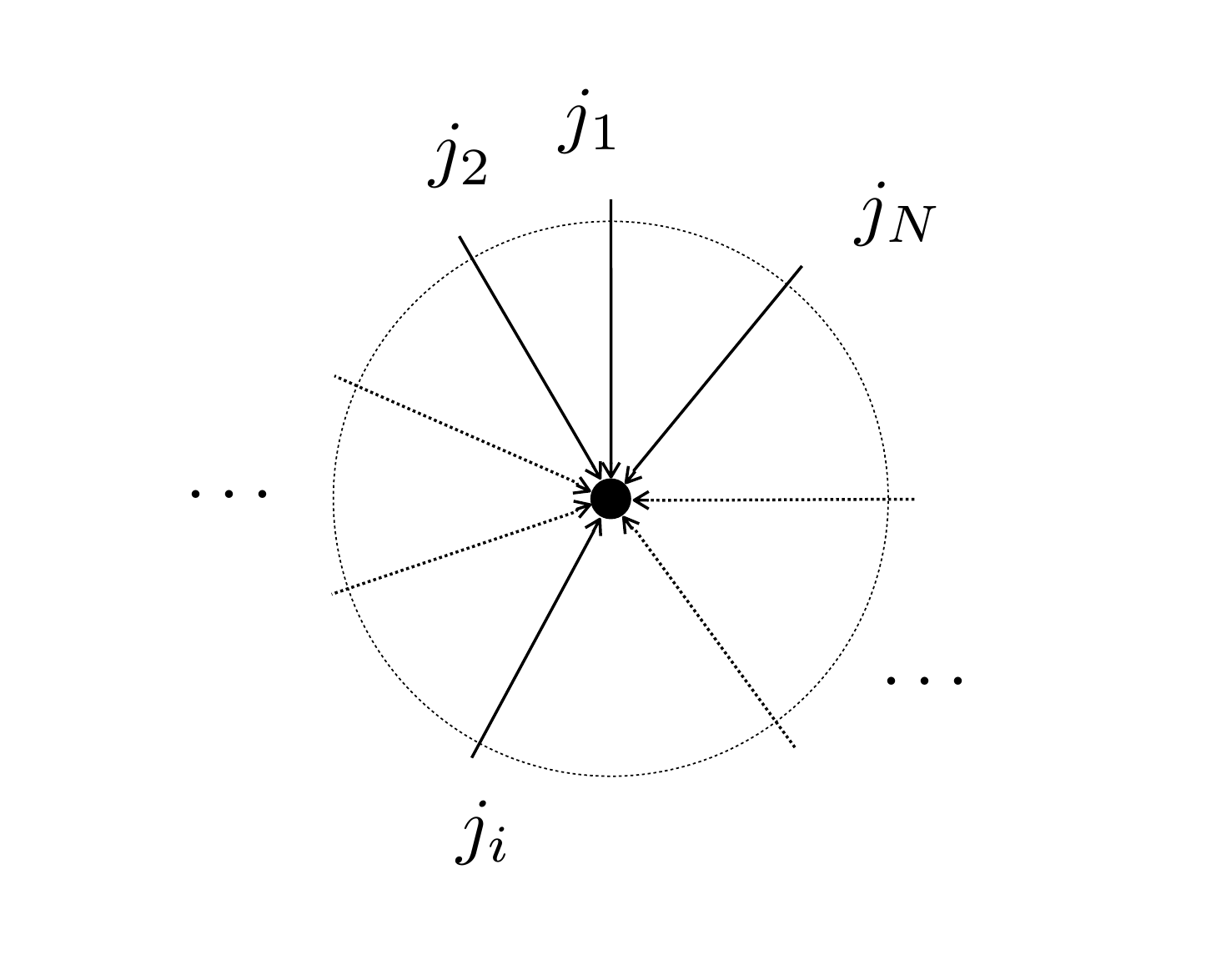}
\caption{The $N$-legged intertwiner system describes a convex polyhedron with $N$ faces, with the topology of a \emph{2-sphere}. The $N$ (oriented) edges are dual to the elementary $N$ surfaces comprising the surface of the polyhedron \cite{don, Poly1,Poly2,Poly3,Poly4}. The intertwiner contains information on how the elementary surfaces, dual to the links, are combined together to form a surface boundary of the space region dual to the node \cite{thi, 5}.}\label{sphere}
\end{figure}

It has been proven in \cite{fr} that the Hilbert space of the $N$-valent intertwiners naturally decomposes into subspaces of constant total area\footnote{The choice of a linear area spectrum $j \times l_P^2$ is favoured by the forthcoming approach involving the $U(N)$ structure of the intertwiner space.} which, following the notation in \cite{fr} we call $\mathcal{H}_N$. Therefore $\mathcal{H} = \bigoplus_{J} \mathcal{H}_N^{(J)}$. We further constrain our system by considering only the invariant \emph{tensor product} Hilbert space, with total spin fixed to $J=J_0$ (see Fig. \ref{sphere}). This is the last ingredient. Eventually, the constrained Hilbert space is given by $\mathcal{H}_{\mathcal{R}} = \mathcal{H}_N^{(J_0)}$. \\

It was also proven in \cite{fr} that each subspace $\mathcal{H}_N^{(J)}$ of $N$-valent intertwiners with fixed total area $J$ carries an irreducible representation of $U(N)$. In this context, one can interpret $J_0$ as the total area dual to the set of $N$ legs of the intertwiner. The main reason behind the choice of $\mathcal{H}_N^{(J_0)}$ as constrained space is that in the semiclassical limit one can think of this system, dually, as a closed surface with area $J_0 l^2_P \gg l^2_P$, where $l_P$ is the Planck length. \\


\subsection{The canonical states of the system}\label{redu}

Once the constrained space has been defined, in order to compute the canonical reduced state, we need the expression of the maximally mixed state $\mathcal{I}_{\mathcal{R}}$ over $\mathcal{H}_{\mathcal{R}}$. This is formally given by

\begin{align}
&\mathcal{I}_{\mathcal{R}} \equiv \frac{1}{d_{\mathcal{R}}}{\mathbb{1}_{\mathcal{R}}} = \frac{1}{d_{\mathcal{R}}} P_{\mathcal{R}} ,
\end{align}

where $P_{\mathcal{R}}$ projects the states of $\otimes_{l} \mathcal{H}^{j_l}$ onto the $SU(2)$ gauge invariant subspace with fixed total spin number $\mathcal{H}_N^{(J_0)}$.

When dealing with $SU(2)$ quantum numbers there are two common choices for the basis of the Hilbert space: the coupled and the decoupled basis. The coefficients which connect the two basis are the well-known Clebsh-Gordan coefficient. Considering that the main task is to perform the partial trace of $\mathcal{I}_{\mathcal{R}}$ over the environment, a suitable basis to write the projector $P_{\mathcal{R}}$ is a \emph{semi-decoupled} basis in which all the quantum numbers within the system and within the environment, respectively, are coupled, but the environment and the system are not.\\

\begin{figure}[h!]
\includegraphics[width=3 in]{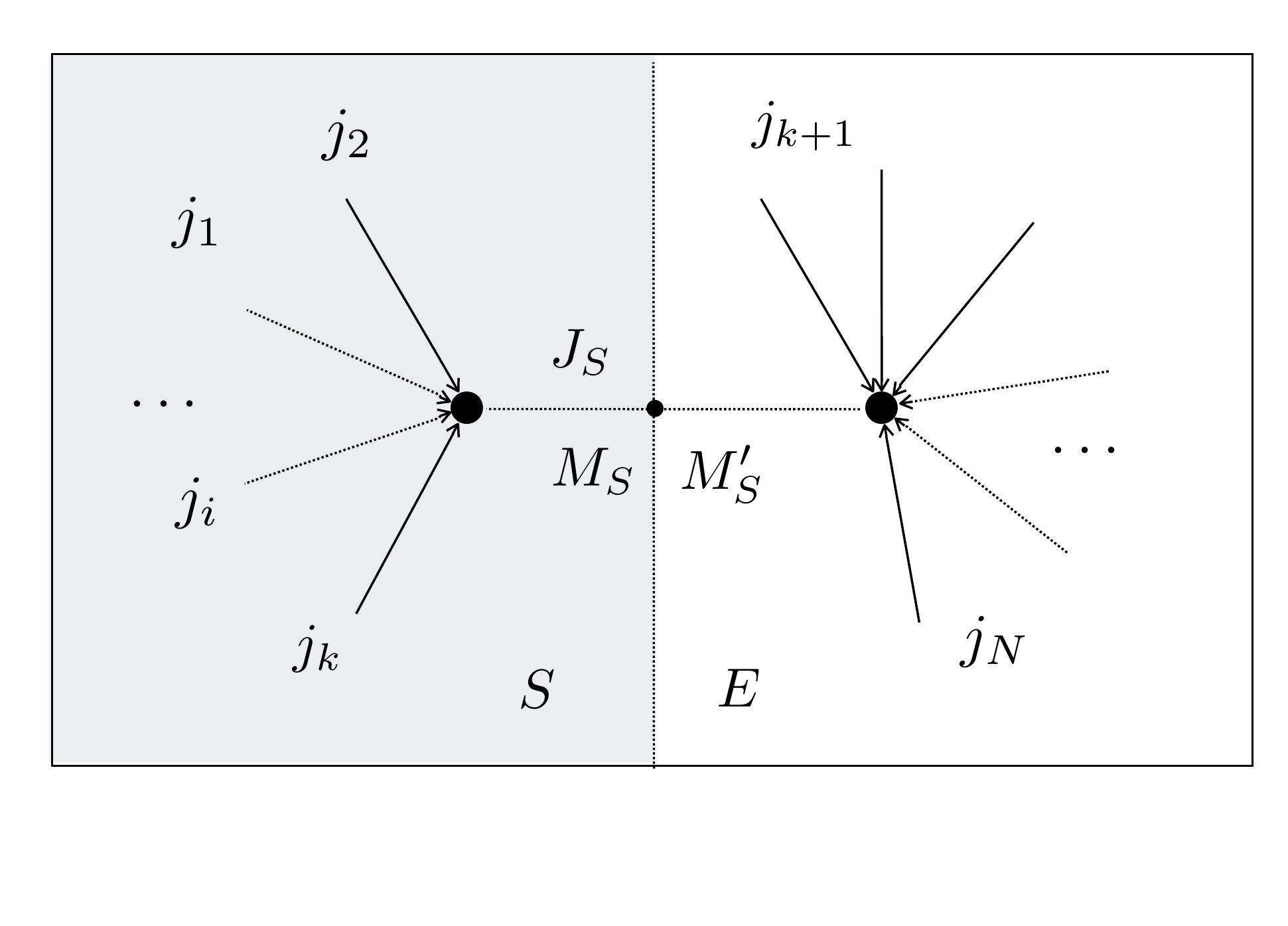}
\caption{Here we show a graphic illustration of the semi-decoupled basis that we are using to write the projector $P_{\mathcal{R}}$ onto the constrained Hilbert space $\mathcal{H}_{\mathcal{R}}$. For a recent work on the splitting of a gauge-invariant system we suggest \cite{Split}.}
\end{figure}

Using such semi-decoupled basis we can write the projector as

\begin{align} \nonumber
&P_{\mathcal{R}} = \sum^{(J_0)}_{\{j_E, j_S\}} \sum_{\eta_E, \sigma_S} \sum_{|\vec{J}_S|,M_S,{M'_S}} \frac{(-1)^{M_S + M'_S}}{d_{|\vec{J}_S|}} \cdot \nonumber \\
& \quad \cdot \Ket{\{j_E, j_S\}; \eta_E, \sigma_S; |\vec{J}_S|, -M_S; |\vec{J}_S|, M_S} \\ \nonumber
& \qquad  \qquad \Bra{\{j_E, j_S\}; \eta_E, \sigma_S; |\vec{J}_S|, -M'_S; |\vec{J}_S|, M'_S} 
\end{align}

where with $\Ket{\{j_E, j_S\}; \eta_E, \sigma_S; |\vec{J}_S|, -M_S; |\vec{J}_S|, M_S}$ we mean $\Ket{\{j_E\} ; \eta_E; |\vec{J}_S|, -M_S}_E  \otimes \Ket{ \{j_S\}; \sigma_S; |\vec{J}_S|, M_S }_S$, $d_{|\vec{J}_S|} \equiv 2 |\vec{J}_S| +1$ and $\sum^{(J_0)}_{\{j_E, j_S\}}$ means that we are summing only over the configurations of the spins $\left\{ j_i\right\}$ such that $\sum_{i \in E} j_i + \sum_{k \in S} j_k = J_0$. The quantum numbers  $\sigma_S$ and $\eta_E$ stand for the recoupling quantum numbers necessary to write the state in the coupled basis, respectively within the system and the environment. Eventually, $|\vec{J}_S|$ and $M_S$ are, respectively, the norm of the total angular momentum of the system and its projection over the $z$ axis; $|\vec{J}_E|$ and $M_E$ have the same meaning but they refer to the environment.

The details of the generic element of the semi-decoupled basis and of the way in which we obtain the projector can be found in the Supplementary Material.\\

The dimension of the  constrained Hilbert space $d_{\mathcal{R}} \equiv \mathrm{dim} (\mathcal{H}_{\mathcal{R}}) $ counts the degeneracy of the $N$-valent intertwiners with fixed total spin $J_0$. Given the equivalence between the space $\mathcal{H}_N^{(J_0)}$ of $N$-valent intertwiners with fixed total area $\sum_i j_i=J_0$ (including the possibility of trivial SU(2) irreps) and the irreducible representation of $U(N)$ formalism for $SU(2)$ intertwiners \cite{fr}, $d_{\mathcal{R}}$ can be calculated as the dimension of the equivalent maximum weight $U(N)$ irrep with Young tableaux given by two horizontal lines with equal number of cases $J_0$,
\begin{align} \label{dime}
d_{\mathcal{R}}=  \frac{1}{J_0+1} \binom{N + J_0 -1}{J_0} \binom{N+ J_0 -2}{J_0} 
\end{align}

 Thanks to the tensor product structure of the semi-decoupled basis, with respect to the bipartition of the universe into system and environment, we can easily perform the partial trace operation over the environment. The details of the computation can be found in the Supplementary Material. The final expression of the canonical state of the system is
 
\begin{align}
&\Omega_S = \sum_{J_S\le J_0/2}  \sum_{\sigma_S, |\vec{J}_S|, M_S} \sum_{\{{j}_S \}}^{J_S} \frac{D_{(N-k)}( |\vec{J}_S|, J_0-J_S )}{d_{|\vec{J}_S|}\,d_{\mathcal{R}}} \cdot \\ 
& \quad \cdot \Ket{\{j_S\}, \sigma_S,|\vec{J}_S|,M_S}  \Bra{\{j_S\},\sigma_S,|\vec{J}_S|,M_S} \nonumber \label{canonico}
\end{align}
\begin{figure}[t]
\includegraphics[width=2.5 in]{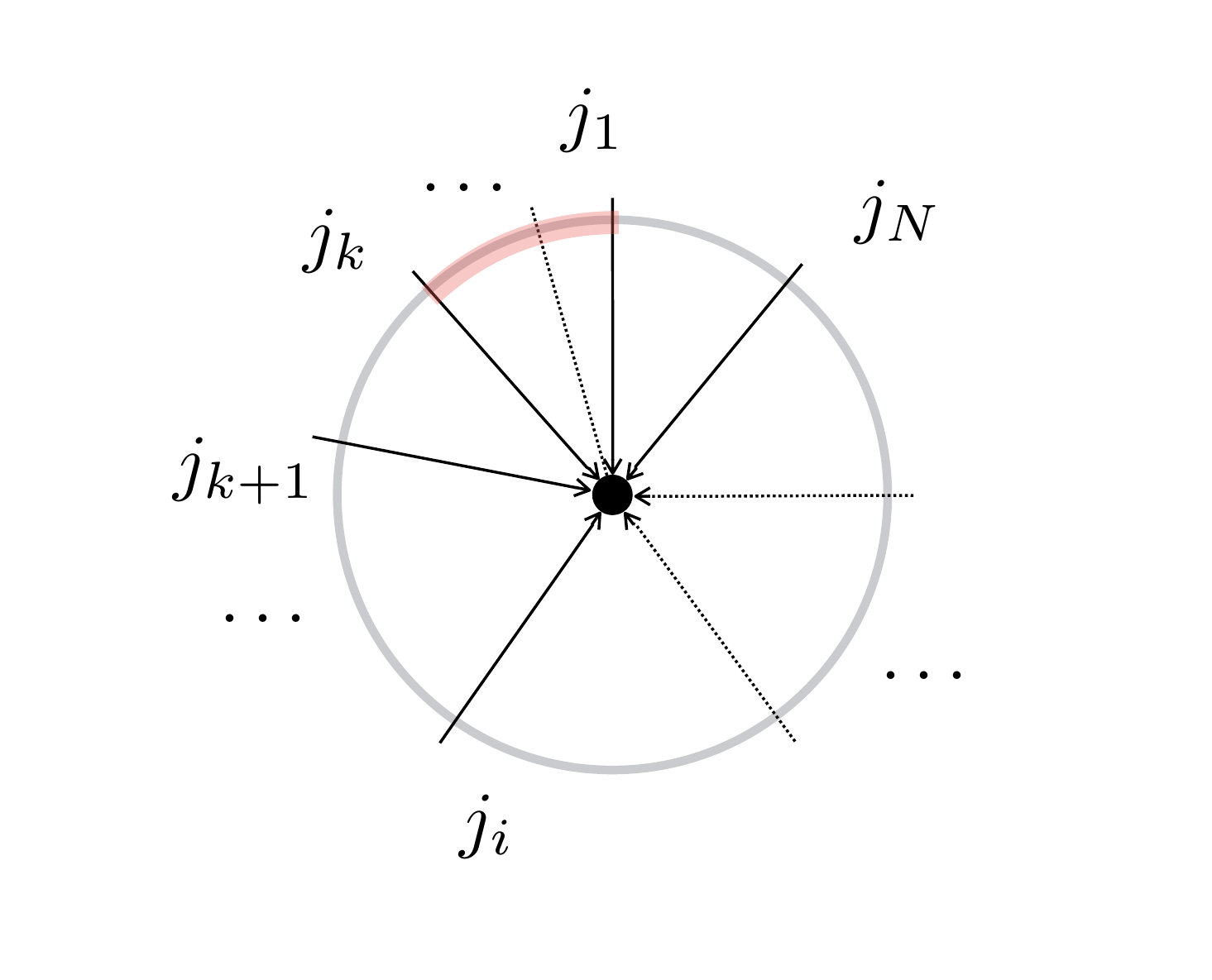}
\caption{A local patch of the 2d surface (in red), associated to a subset of intertwined links $\{ j_{1},\cdots ,j_k\}$ defining the ``system''. The ``environment'' is identified with the complementary 2d-surface associated to the set of links  $\{ j_{k+1},\cdots ,j_N\}$, with $N \gg k$.}\label{sphere2}
\end{figure}

Where $\sum_{\{{j}_S \}}^{J_S}$ means that we are summing over the configurations of the spins of the system $\left\{ j_S\right\}$ such that $\sum_{k \in S} j_k = J_S$. Moreover the definition of the $D-$functions is

\begin{align} 
&D_{(Q)}( x,y )\equiv \\ \label{canoni}
& \qquad \frac{2x + 1}{x+y + 1} \binom{Q + y + x - 1}{x+y} \binom{Q + y - x - 2}{y - x} \nonumber 
\end{align}


We also define the following short-hand notation $W_{\mathcal{E}} \equiv D_{(N-k)}( |\vec{J}_S|, J_0-J_S )$. We will also call $W_{\mathcal{S}}\equiv D_{(k)}( |\vec{J}_S|, J_S )$ the dimension of the system's degeneracy space with fixed area $J_S$ and closure defect $ |\vec{J}_S|$, derived from the equivalent $U(N)$ representation as for the case of the environment in \eqref{canoni}. 



The canonical weight $W_{\mathcal{E}}$ encodes all the information about the local structure of correlations of the reduced intertwiner state. The specific form of this factor tells us about the physics of the system, defined by the specific choice of constraints: the SU(2) gauge symmetry and the fixed total area constraint. Given the global constraint, the split in system and environment breaks the gauge symmetry. Due to the presence of the constraint, onto $\mathcal{H}_{\mathcal{R}}$ the quantum numbers of system $\left( \left\{ j_S \right\}, \sigma_S, |\vec{J}_S|, M_S \right)$ are intertwined with those of the environment $\left( \left\{ j_E \right\}, \eta_E, |\vec{J}_{E}| =|\vec{J}_{S}|, M_E = - M_S \right)$. This is why, beside the expected dependence on the total area of the system $J_S$, the canonical weight carries some interesting extra information on the local closure defect $ |\vec{J}_S|$.  

\section{Typicality of the reduced state}\label{typ}

In this section we study the region of the space of the parameters $(N,k,J_0,J_{max})$ where the canonical reduced state is typical. In other words, we investigate the distance of the canonical state from a randomly chosen pure state in $\mathcal{H}_{\mathcal{R}}$. 

Concretely, following the approach described in Section \ref{tysn}, we want to show that for the overwhelming majority of intertwiner states $|\mathcal{I}\rangle \in \mathcal{H}_{\mathcal{R}} \subseteq \mathcal{H}_E \otimes \mathcal{H}_S$, the trace distance $D(\rho_S,\Omega_S)$ between the reduced density matrix of the system $\rho_S = Tr_E(|\mathcal{I} \rangle \langle \mathcal{I}|)$ and the canonical state $\Omega_S$ is extremely small\footnote{We remember that the trace-distance has an important physical interpretation: $D(\rho,\sigma)$ is the probability two tell apart $\rho$ and $\sigma$, by means of the most effective quantum measurement\cite{Nielsen}}. This amounts to prove two things: first, that the Hilbert space average of such trace distance is itself quite small in the regime in which we are interested in
\begin{align}
&\mathbb{E} \left[ D(\rho_S,\Omega_S)\right] \ll 1 \,\,,
\end{align}

where $\mathbb{E}$ indicates the Hilbert space average performed using the unique unitarily invariant Haar measure \cite{3, Geos}. Second, that the fraction of states for which such distance is higher than a certain $\epsilon$ is exponentially vanishing in the dimension of the Hilbert space. \\

Now, following \cite{popescu}, one can recast the condition in \eqref{result} with the following bound on the average distance,
\begin{align} \label{bbound}
&0 \leq \mathbb{E} \left[ D(\rho_S,\Omega_S)\right] \leq \sqrt{\frac{d_S}{d_E^{\mathrm{eff}}}} \leq \frac{d_S}{\sqrt{d_{\mathcal{R}}}}
\end{align}

Concretely, the first step toward the statement of typicality in our context amounts to study in which region of the parameters space $(J_0,N,k,J_{max})$ we have $d_S/\sqrt{d_{\mathcal{R}}} \ll 1$.

\subsection{Evaluation of the bound}

The Hilbert space of the system is the tensor product Hilbert space of the set of irreps $V^{j_i}$ with a given cutoff $J_{max}$. We assume $J_{max}\ge J_0$, in order to be sure that $\mathcal{H}_{N}^{(J_0)}$ will always carry an irreducible representation of $U(N)$.  Each $V^{j}$ has dimension $d_j=2j+1$. Therefore, considering the set of $k$ edges comprising the system, we have
\begin{align}
d_S = \prod_{i=1}^{k} \sum_{j_i=0,\frac{1}{2}}^{J_{max}}(2j_i+1) = \left( 2J_{max}+1\right)^{k} \left( J_{max}+1\right)^{k}
\end{align}

Analogously, for the environment we get $d_E = \left( 2J_{max}+1\right)^{N-k} \left( J_{max}+1\right)^{N-k}$. 

Since $d_{\mathcal{R}}$ is given in \eqref{dime}, we can focus on the last inequality in \eqref{bbound} and define the regime where $\mathbb{E} \left[ D(\rho_S,\Omega_S)\right] \ll 1 $. 

Studying the ratio
\begin{align}
&\frac{d_S^2}{d_{\mathcal{R}}} = \frac{(2 J_{\mathrm{max}} + 1)^{2k}(J_{\mathrm{max}} +1 )^{2k}}{\frac{1}{J_0 + 1} \binom{N+J_0 -1}{J_0} \binom{N+J_0 -2}{J_0}} \label{ratio}
\end{align}

we can see that $N$ and $J_0$ play a rather symmetric role in making this quantity small. The region of interest is certainly $J_0 \gg1$ or $ N \gg 1$, or both. As we will argue in the next section, $J_0,N \gg 1$ is precisely the regime of interest for the thermodynamical limit. Therefore we focus on this region, where there are two different regimes: $J_0 \gg N \gg 1$ or $N \geq J_0 \gg 1$. In both cases there are wide regions of the parameters space where the inequality $\mathbb{E} \left[ D(\rho_S,\Omega_S)\right] \ll 1$ holds. We were able to extract the following two conditions which guarantee an exponential decay of \eqref{ratio}, either on $N$ or on $J_0$:

\begin{subequations}
\begin{align}
&\frac{J_0}{k} > \log J_{max} && (J_0 \gg N \gg  1) \label{cut1}\\
&\frac{N}{k} \log j_0 > 2 \log J_{max} &&  (N \geq J_0 \gg  1) \label{cut2}
\end{align}
\end{subequations}

The details can be found in the supplementary material but we would like to present a physically motivated argument to provide a meaningful value for the cut-off $J_{max}$ and check the plausibility of the given bounds. As argued in \cite{CarloEugenio}, if we look at a sphere with small radius $l$, placed at a large distance $L$, we will see it within a small angle $\phi \sim \frac{l}{L}$. Therefore using the scale of the radius of the observed universe $L_U$ and assuming that there is nothing with size smaller than the planck length $l_P$, we will never see something with angular extension smaller than $\phi_{min} \sim \frac{l_P}{L_U}$.\\

A spherical harmonics of representation $j$ is able to discriminate angular distances of the order $\frac{4\pi}{2j+1}$. Therefore the existence of $\phi_{min}$ means that there is an upper bound to the representation which we need to consider which is $J_{max} \sim \frac{4\pi}{\phi_{min}^2}  = 4\pi \frac{L_U^2}{l_P^2}$. Using this argument we obtain the following cut-off
\begin{align}
&J_{max} \sim 4\pi \frac{L_U^2}{l_P^2} \approx 3 \times 10^{124} \sim e^{124 \times \log 10}
\end{align}

Putting the numbers in \eqref{cut1} and \eqref{cut2} we obtain

\begin{subequations}
\begin{align}
& \frac{J_0}{k} \gtrsim 3 \times 10^2 && (J_0 \gg N \gg  1)\\
& \frac{N}{k} \gtrsim 6 \times 10^2 && (N \geq J_0 \gg  1)
\end{align}
\end{subequations}

\subsection{Levy's lemma}

Following \cite{popescu,popescu2}, we can use Levy's lemma (see Appendix \ref{levy}) to bound the fraction of the volume of states which are $\varepsilon$ more distant than $\frac{d_S}{\sqrt{d_{\mathcal{R}}}}$ from $\Omega_S$ as

\begin{align} 
&\frac{\mathrm{Vol} \left[  |\mathcal{I}\rangle \in \mathcal{H}_{\mathcal{R}}\, \vert \, D(\rho_S,\Omega_S) - \frac{d_S}{\sqrt{d_{\mathcal{R}}}}\geq \varepsilon \right] }{\mathrm{Vol} \left[|\mathcal{I}\rangle \in \mathcal{H}_{\mathcal{R}} \right] } \leq B_\epsilon(d_{\mathcal{R}})\\ \nonumber
& B_{\epsilon} (d_{\mathcal{R}})\equiv 4 \, \mathrm{Exp} \left[-\frac{2}{9\pi^3} d_{\mathcal{R}} \varepsilon^2 \right].
\end{align}

The dimension $d_{\mathcal{R}}$ can be evaluated numerically because we have an exact expression. We give a numeric example to show that it is not necessary to have huge areas or number of links for the typicality to emerge. Suppose we have a huge sensitivity on the trace distance: $\epsilon = 10^{-10}$. Moreover, $\frac{2}{9\pi^3} \sim  7 \times 10^{-3}$. With these numbers we have

\begin{align}
&B_{10^{-10}}(d_{\mathcal{R}}) = 4\mathrm{Exp} \left[-7\cdot 10^{-23} d_{\mathcal{R}}  \right].
\end{align}

Suppose we look at the most elementary patch, just a few links ($k=1,2$). The set of numbers $J_0 = N = 10^4$ gives the following bounds, using a cut-off given by the cosmological horizon 

\begin{subequations}
\begin{align}
&\frac{N}{k} \sim 10^4 \gg 6 \times 10^2 \\
& B_{10^{-10}}(d_{\mathcal{R}}) = 4\mathrm{Exp} \left[-5.6 \times 10^{5992}  \right] \ll 1
\end{align}
\end{subequations}

As we can see, the typicality emerges quite easily, due to the exponential-like growth of the constrained Hilbert space on the number of links $N$ and on the total area $J_0$.\\

The existence of a typical behaviour indicates the emergence of a regime where the properties of the reduced state of the $N$-valent intertwiner state are \emph{universal}. The structure of \emph{local} correlations carried by the reduced state is independent from the specific shape of the pure intertwiner state and it is locally the same everywhere. Due to the global symmetry constraint though, the canonical weight presents a very involved analytic form, despite the extreme simplicity of the system under study. In order to extract some physical information from this coefficient we are going to study its behaviour in the thermodynamic limit. 

\section{Thermodynamic limit \& area laws}\label{thermo}

In this section we investigate the behaviour of the entropy of the reduced state, in the thermodynamic limit. 

In the standard context of statistical mechanics, when performing the thermodynamic limit the density of particles must be finite otherwise the energy density would diverge: $N,V \to + \infty$ with $\frac{N}{V} < + \infty$. As we will see in the forthcoming argument, the area is playing here the role of the energy, therefore we think that the correct way of performing the thermodynamic limit consists in taking $N,J_0 \to \infty$ with $\frac{J_0}{N} \equiv j_{0}  < +\infty$, where $j_0$ is the average spin of the intertwiner.\\

The entropy of the system is given by the von Neumann entropy,
\begin{align}
&S(\Omega_S)=-\text{Tr}[\Omega_S\,\log \Omega_S]
\end{align}
Given the diagonal form of the canonical reduced density matrix $\Omega_S$ in \eqref{canonico}, this can be written as
\begin{align} \label{entro}
&S(\Omega_S)= -  \frac{1}{d_{\mathcal{R}}} \sum_{J_S\le J_0/2, |\vec{J}_S|} W_{\mathcal{S}} \,W_{\mathcal{E}}\log{\left(\frac{W_{\mathcal{E}}}{d_{|\vec{J}_S|}\,d_R}\right)}.
\end{align}

Within the typicality regime ($N, J_0 \gg1$) we can use the Stirling approximation for the factorials, to simplify the form of the binomial coefficients in $W_{\mathcal{E}}, W_{\mathcal{S}}$ and $d_{\mathcal{R}}$. We will study separately the three regimes $j_0 \gg 1$, $j_0 \ll 1$ and $j_0 \sim 1$. The details of the computation can be found in the supplementary material, here we only summarise the results.

\subsubsection{Small average spin: $j_0 \ll 1$}

In the case of small average spin, the leading term in the thermodynamic limit is 

\begin{align} \label{area}
&S(\Omega_S) \simeq \beta \langle 2J_S\rangle+\text{small corrections}\end{align}
where $\langle \cdot \rangle$ is the quantum mechanical average, on the canonical state $\Omega_S$, while
\begin{align} 
\beta \equiv \left(1+\log{\frac{N-k}{J_0}} \right)
\end{align}
is formally identified as the ``temperature'' of the environment. It turns out to be a function of the averaged spin of the environment.

Despite being quite far from the standard setting, a hint toward a thermodynamical interpretation of this result comes from the $U(N)$ description of the $SU(2)$ intertwiner space. Using the Schwinger representation of the $\mathfrak{su}(2)$ Lie algebra \cite{gir, fr}, one can describe the $N$-valent intertwiner state as a set of $2N$ oscillators, $a_i, b_j$. The quadratic operators $E_{ij}\equiv(a^{\dagger}_i a_j- b^{\dagger}_i b_j), E^{\dagger}_{ij}= E_{ji} $ acts on couples of punctures $(i, j)$ and form a closed $\mathfrak{u}(N)$ Lie algebra. The $\mathfrak{u}(1)$ Casimir operator is given by the oscillators' energy operator $E\equiv \sum_i E_i$, with $E_i \equiv E_{ii}$, and its value on a state gives twice the sum of the spins on all legs, $2 \sum_i j_i$. Therefore, one can interpret $E$ as measuring (twice) the total area of the boundary surface around the intertwiner.\\

In statistical mechanics, the thermal behaviour of the canonical state relies on the constraint of energy conservation. The emergence of the canonical state from the micro-canonical occurs as the degeneracy of the environment grows exponentially with the energy, hence decreasing exponentially with the system energy.

In these terms, constraining the total area is equivalent to fix a shell of eigenvalues (in fact a single eigenvalue) of the energy operator acting on the full system. In the limit $N\gg J_0 \gg 1$, the degeneracy of the single energy level grows exponentially.

For such a reason the area scaling described by \eqref{area} is consistent with a thermal interpretation for our reduced surface state. It is also worth to mention that the departure from the exact thermal behaviour, \`{a} la Gibbs, is a signature of the breaking of the global SU(2) symmetry (closure defect), witnessed by the explicit dependence of the reduced state on $|\vec{J}_S|$. 

\subsubsection{High average spin: $j_0 \gg 1$}

Here we study the behaviour of the entropy in the regime $J_0\gg N \gg 1$. Up to $O(1/J_0)$ the logarithm of the normalised canonical weight is given by
\begin{align} \label{mah}
&-\log{\left(\frac{W_{\mathcal{E}}}{d_{J_S}\,d_R}\right)} \simeq -\log \left( \frac{J_0e}{N-k} \right)^{-2k} + \frac{3k}{N}-\frac{2kJ_S}{J_0}\\ \nonumber
&- \frac{2J_S+2|\vec{J}_E|}{J_0} \simeq k \log \left( \frac{J_0e}{N-k} \right)^2 + \text{small corrs}
\end{align}

Interestingly, the leading term does not depend on the quantum numbers of the system. Therefore the entropy is counting the number of orthogonal states on which the canonical state has non-zero support

\begin{align}
&S(\Omega_S) \simeq 2k \left( 1+ \log \left( \frac{J_0}{N} \right) \right)+ O\left(\frac{k}{N}, \frac{1}{J_0}\right) \,\,\, . \label{eq:ent2}
\end{align}
This makes the entropy extensive in the number of edges comprising the dual surface of the system. In this sense, the term $\left[ \left( \frac{J_0e}{N-k} \right)^2\right]^k$ defines some kind of \emph{effective} dimension of the system, suggesting that the following two things happen in such regime: first, the canonical state has approximately a tensor product structure; second that the total spin is equally distributed among all spins in the universe therefore the accessible Hilbert space of each spins is roughly limited by a representation of the order of $j_0$. 
The validity of this interpretation can be checked assuming a tensor product structure of $k$ links with single-link Hilbert space limited to the representation $\alpha \times j_0$ and computing the entropy $S_{eff}$ as the logarithm of the dimension of this space. If we can find an $\alpha \sim 1$ such that the difference $S(\Omega_S) - S_{eff}$ is proportional only to small corrections $O(\frac{1}{N},\frac{1}{J_0})$, we can say that our argument is not too far from what is happening in such a regime. With these assumptions the effective dimension of the Hilbert space of the system is 

\begin{align}
&d_S^{eff} =\prod_i  \sum_{j_i = 0, \frac{1}{2}}^{\alpha j_0}  (2j_i +1) = (2 \alpha j_0 + 1)^{k}(\alpha j_0 + 1)^k 
\end{align}

In the $j_0 \gg 1$ regime we can write it as $d_S^{eff} \simeq 2^k \alpha^{2k} j_0^{2k} + O(\frac{k}{j_0})$ which gives

\begin{align}
&S_{eff} \equiv \log d_S^{eff} \simeq 2k \log \left( \frac{J_0}{N} \right) + k \left(\log 2 \alpha^2 \right)
\end{align}

The difference between the two entropies 

\begin{align}
&S(\Omega_S) - S_{eff} \simeq k(2-\log 2\alpha^2 ) + O\left( \frac{1}{N}, \frac{1}{J_0}\right)
\end{align}

is given only by small corrections of order $O\left( \frac{1}{N}, \frac{1}{J_0}\right)$ when $\alpha \simeq 2$. 

This simple computation provides evidence that the result in Eq.(\ref{eq:ent2}) follows from the two aforementioned assumptions.

\subsubsection{Order $1$ average spin: $j_0 \sim 1$}

Eventually, we compute the behaviour of the entropy in the intermediate regime $J_0 \sim N \gg 1$. With respect to the previous cases, this regime does not add anything new to the analysis. The observed behaviour is  extensive in the number of links of the system, with a coefficient which is slightly different from the previous one:

\begin{align}
&S(\Omega_S) \simeq \left(2k-3\right) (1+\mathcal{O}\left[ \left( \frac{k}{N}\right)^2\right])
\end{align}

The relevant computation can be found in the supplementary material.

\section{Summary and Discussion}\label{fine}

In this manuscript we extend the so-called typicality approach, originally formulated in statistical mechanics contexts, to a specific class of tensor network states given by $SU(2)$ invariant spin networks. In particular, following the approach given in \cite{popescu}, we investigate the notion of canonical typicality for a simple class of spin network states given by $N$-valent intertwiner graphs with fixed total area.  Our results do not depend on the physical interpretation of the spin-network, however they are mainly motivated by the fact that spin networks provide a gauge-invariant basis for the kinematical Hilbert space of several background independent quantum gravity approaches, including loop quantum gravity, spin-foam gravity and group field theories.

The first result is the very existence of a regime in which we show the emergence of a canonical typical state, of which we give the explicit form. Geometrically, such a reduced state describes a patch of the surface comprising the volume dual to the intertwiner. The structure of correlations described by the state should tell us how local patches glue together to form a closed connected surface in the quantum regime. 

We find that, within the typicality regime, the canonical state tends to an exponential of the total spin of the subsystem with an interesting departure from the Gibbs state. 
The exponential decay \`a la Gibbs of the reduced state is perturbed by a parametric dependence on the norm of the total angular momentum vector of the subsystem (closure defect).  Such a feature provides a signature of the non local correlations enforced by the global gauge symmetry constraint. This is our second result. 

We study some interesting properties of the typical state within two complementary regimes, $N \gg J_0 \gg 1$ and $J_0 \geq N  \gg 1$. In both cases, we find that the area-law for the entropy of a surface patch must be satisfied at the local level, up to sub-leading logarithmic corrections due to the unavoidable dependence of the state from the closure defect. However, the area scaling interpretation of the entropy in the two regimes is quite different. In the $N \gg J_0 \gg 1$ regime, the result is related to the definition of a generalised Gibbs equilibrium state. The area is playing the role of the energy, as imposed by the specific choice of the global constraint, requiring total area conservation.  

On the other hand, in the $J_0 \geq  N  \gg 1$ regime, the area scaling is given by the extensivity of the entropy in the number of links comprising the reduced state, as for the case of the generalised (non $SU(2)$-gauge invariant) spin networks \cite{donne}. In this regimen, each link contributes independently to the result, indicating that the global constraints are very little affecting the local structure of correlations of the spin network state. Still, interestingly, the remainder of the presence of the constraints can be read in the definition of what looks like an effective dimension for the single link Hilbert space. 

We interpret these results as the proof that, within the typicality regime, there are certain (local) properties of quantum geometry which are ``universal'', namely independent of the specific form of the global pure spin network state and descending directly from the physical definition of the system encoded in the choice of the global constraints. 

We would like to stress that our result is purely kinematic, being a statistical analysis on the Hilbert space of spin-network states. For the case of a simple intertwiner state, such study necessarily requires to consider a system with a \emph{large number} of edges, beyond the very large dimensionality of the Hilbert space of the single constituents. 
In this sense, the presented statistical analysis and thermal interpretation is very different from what recently done in \cite{chi, chi1,ha}, considering quantum geometry states characterised by few constituents with a high dimensional Hilbert space. 
In fact, we expect a {large number} statistical analysis to play a prominent role in facing the problem of the continuum in quantum gravity. Therefore we think it is important to propose and develop new technical tools which are able to deal with a large numbers of elementary constituents and extract physically interesting behaviours.\\

The kinematic nature of the statement of typicality, together with its general formulation in terms of constrained Hilbert spaces given in \cite{popescu}, provide an important tool to study the possibility of a thermal characterisation of reduced states of quantum geometry, regardless of any hamiltonian evolution in time. Beyond the simple case considered in the paper and in a more general perspective, we expect typicality to be useful to understand how large the effective Hilbert space of the theory can be, given the complete set of constraints defining it. It will also help in understanding which typical features we should expect to characterise a state in such space. If we think of dynamics as a flow on the constrained Hilbert space, we generally expect that, even if the initial state is highly un-typical after a certain ``time''-scale we will find the system in a state which is extremely close to the typical state. This happens because, as it has been shown in the original paper on typicality, the number of states close to the typical state are the overwhelming majority.\\

Finally, it is interesting to look at the proposed ``generalised'' thermal characterisation of a local surface patch, within the standard LQG description of the horizon, as a closed surface made of patches of quantized area. Differently from the \emph{isolated horizon} analysis (see e.g. \cite{pra, alex, ori1}), in our description the thermal character of the local patch is not (semi)classically induced by the thermal properties of a black hole horizon geometry, but emerges from a purely quantum description. In this sense, our picture goes along with the informational theoretic characterisation of the horizon proposed in \cite{ter}. 

In fact, we think that typicality could be used to define an information theoretic notion of quantum horizon, as the boundary of a generic region of the quantum space with an emergent thermal behaviour. We leave this for future work.




\subsection*{Acknowledgements}

The authors are grateful to Daniele Oriti, Aldo Riello and Thibaut Josset for interesting discussions and careful readings of the draft of the paper.

\appendix

\section{The Levy-lemma} \label{levy}
In order to better understand the result it is useful to look at its most important step, which is the so-called Levy-lemma. Take an hypersphere in $d$ dimensions $S^{d}$, with surface area $V$. Any function $f$ of the point which does not vary too much 

\begin{align}\nonumber
&f :  S^d \ni \phi \to f(\phi) \in \mathbb{R} &&|\nabla f| \leq 1
\end{align}

will have the property that its value on a randomly chosen point $\phi$ will approximately be close to the mean value.
\begin{align}\nonumber
\frac{\mathrm{Vol}\left[ \phi \in S^d \, : \, f(\phi) - \MV{f} \geq \epsilon \right]}{\mathrm{Vol}\left[ \phi \in  S^d \right]} \leq 4 \, \mathrm{Exp} \left[ - \frac{d+1}{9 \pi^3} \epsilon^2 \right]
\end{align}

Where $\mathrm{Vol}\left[ \phi \in S^d \, : \, f(\phi) - \MV{f} \geq \epsilon \right]$ stands for ``the volume of states $\phi$ such that $f(\phi) - \MV{f} \geq \epsilon$''. $\MV{f}$ is the average of the function $f$ over the whole Hilbert space and $\mathrm{Vol}\left[ \phi \in  S^d \right]$ is the total volume of the Hilbert space. Integrals over the Hilbert space are performed using the unique unitarily invariant Haar measure.\\

The Levy lemma is essentially needed to conclude that all but an exponentially small fraction of all states are quite close to the canonical state. This is a very specific manifestation of a general phenomenon called ``concentration of measure'', which occurs in high-dimensional statistical spaces \cite{led}.

The effect of such result is that we can re-think about the ``a priori equal probability'' principle as an ``apparently equal probability'' stating that: as far as a small system is concerned almost every state of the universe seems similar to its average state, which is the maximally mixed state $\mathcal{E}_{\mathcal{R}} = \frac{1}{d_{\mathcal{R}}}\mathbb{I}_{\mathcal{R}}$.\\

\end{document}